\documentclass[fleqn,usenatbib]{mnras}

\usepackage{newtxtext,newtxmath}

\usepackage[T1]{fontenc}

\DeclareRobustCommand{\VAN}[3]{#2}
\let\VANthebibliography\thebibliography
\def\thebibliography{\DeclareRobustCommand{\VAN}[3]{##3}\VANthebibliography}

\usepackage{graphicx}	
\usepackage{amsmath}	
\usepackage[dvipsnames]{xcolor}
\usepackage[normalem]{ulem}
\usepackage{soul}
\usepackage{booktabs}

\newcommand{\sub}[1]{_{\mathrm{#1}}}

\newcommand{\msun}{M\sub{\sun}}

\def\equationautorefname~#1\null{Eq.~(#1)\null}
\def\figureautorefname~#1\null{Fig.~#1\null}
\newcommand{\appref}[1]{\hyperref[#1]{Appendix~\autoref{#1}}}

\title[Stellar distribution of satellite galaxies]{Imprints of tidal interactions on the stellar distribution of satellite galaxies: implications for dark matter deficient galaxies}

\author[Z.H. Yin et al.]{
Zhihao Yin$^{1}$\thanks{E-mail: 12345055@zju.edu.cn (ZY)}, 
Go Ogiya$^{1,2}$\thanks{Email: gogiya@zju.edu.cn (GO), Corresponding author}, 
Frank C. van den Bosch$^{3}$
\\
$^{1}$Institute for Astronomy, School of Physics, Zhejiang University, Hangzhou 310027, China \\
$^{2}$Center for Cosmology and Computational Astrophysics, Zhejiang University, Hangzhou 310027, China \\
$^{3}$Department of Astronomy, Yale University, PO. Box 208101, New Haven, CT 06520-8101
}

\date{Accepted XXX. Received YYY; in original form ZZZ}

\pubyear{\the\year{}}

\begin{document}
\label{firstpage}
\pagerange{\pageref{firstpage}--\pageref{lastpage}}
\maketitle

\begin{abstract}
Interactions with the host galaxy strip stars and dark matter from the outer regions of satellite galaxies. Meanwhile, some stars from the central regions can migrate outward due to dynamical heating, producing an excess in the outer surface brightness relative to the extrapolation of the inner S\'{e}rsic profile. Recently discovered dark matter deficient galaxies (DMDGs) appear to be representative examples of such tidally disturbed systems. In this work, we investigate how the break radius, defined as the radius beyond which this surface brightness excess emerges, forms and evolves, by performing $N$-body simulations of a satellite galaxy interacting with a host, where the satellite serves as a plausible progenitor of a DMDG. Our simulations naturally reproduce a break radius consistent with that observed in DMDGs. We find that the break radius grows over time and exhibits a characteristic evolutionary behaviour: during each pericentric passage it briefly contracts due to tidal compression, and then rapidly and strongly expands as the satellite undergoes dynamical relaxation. After the satellite reaches a quasi-equilibrium configuration, the break radius shows only mild variations until the next pericentre. Across our suite of simulations, the ratio of the break radius to the effective radius remains approximately constant, even when we change the orbital parameters and internal structure of the satellite. Based on these findings, we develop a prescription for predicting the time evolution of the break radius, which can be used to constrain the tidal interaction history of satellite galaxies, including DMDGs and splashback galaxies.
\end{abstract}




\section{Introduction}

The intrinsic properties of satellite galaxies can be substantially transformed by tidal interactions with their host galaxies, making these processes crucial for advancing our understanding of galaxy formation and evolution in the low-mass regime \citep[e.g.,][]{moore_galaxy_1996,bekki_galaxy_2003,boselli_environmental_2006}. Observations of satellite systems have increasingly uncovered clear signatures of such tidal effects encoded in their outer stellar structures. Many satellites exhibit extended stellar structures that deviate from simple equilibrium descriptions, including tidal arms and characteristic S-shaped morphologies \citep[e.g.,][]{majewski_2mass_2003,koch_threshing_2012}, diffuse and extended stellar envelopes \citep{mcconnachie_multiple_2007}, and coherent stellar streams \citep{martinez-delgado_slar_2010,martinez-delgado_dwarfs_2012,sand_tidal_2012}. 

However, the physical origin of these tidal features has so far been addressed mainly in a qualitative manner. With forthcoming facilities capable of deep, wide-field observations, including the Chinese Space Station Telescope \citep{CSST2026}, the Vera C. Rubin Observatory \citep{ivezic_lsst_2019}, and Euclid \citep{scaramella_euclid_2022}, which are anticipated to reveal more detailed outer stellar structures for a larger sample of satellite galaxies, there is an increasing demand for theoretical frameworks that support more quantitative analyses of their dynamical evolution.

Numerical simulations have shown that when a satellite galaxy interacts with its host, material from its outer regions, both stellar and dark matter (DM), is stripped away, and a portion of the stars originally located near the centre of the satellite can migrate outward. These displaced stars generate an excess of surface brightness in the outer parts of the satellite galaxy \citep{aguilar_density_1986, johnston_interpreting_2002, mayer_tidal_2002}, in agreement with those in observations \citep{Choi2002, Muller2017, Tang2020}. This excess leads to a systematic deviation from the S\'{e}rsic profile \citep{sersic_influence_1963} that describes the inner stellar light distribution, thereby introducing a characteristic scale at which this transition occurs. Previous work has proposed that this transition scale, often termed the break radius, can be predicted theoretically \citep{penarrubia_tidal_2008,lokas_tidal_2013}. However, more systematic investigations, along with a definition of the break radius that is more closely matched to observational practice, are still needed.

To investigate the tidal evolution of satellite galaxies, the recently identified dark matter deficient galaxies (DMDGs) near the massive elliptical galaxy NGC\,1052, NGC\,1052-DF2 and NGC\,1052-DF4 \citep[hereafter DF2 and DF4;][]{van_dokkum_galaxy_2018, van_dokkum_second_2019}, represent particularly compelling examples. Within the standard $\Lambda$ Cold Dark Matter ($\Lambda$CDM) framework, it is expected that these galaxies are surrounded by DM haloes with masses that surpass their baryonic mass by a factor of over 300 \citep{behroozi_universemachine_2019}. Nonetheless, the dynamical masses derived from the galactic kinematics reveal a stellar mass-to-halo mass ratio that is approximately unity, suggesting an absence of DM \citep[e.g.,][]{danieli_still_2019,Emsellem2019}. A possible explanation for the origin of DMDGs is that DF2 and DF4 are extreme tidal remnants \citep{ogiya_tidal_2018,Yang2020, maccio_creating_2021,jackson_dark_2021,Moreno2022,katayama_impact_2024}. In this scenario, the progenitors of DMDGs, which start out with ordinary DM haloes, experience strong tidal stripping after they are accreted into a more massive host system, namely the NGC\,1052 group. This stripping process can remove nearly all of their DM, while a large fraction of their stars stays gravitationally bound, thereby transforming normal satellite galaxies into DMDGs.

In this scenario, tidal forces play a dual role by not only stripping away loosely bound DM from the satellite galaxy but also repositioning stars, resulting in the formation of tidal streams and modifying the structure of the stellar body. Recent observations of DF2 and DF4 provide additional evidence supporting this idea \citep[][but see also \citealt{montes_disk_2021}]{montes_galaxy_2020,keim_tidal_2022}: These galaxies exhibit signs of tidal distortion in their stellar distribution, including significant shifts in position angle and elongated isophotes, as well as evidence for a break radius in their outskirts.

Additional scenarios for producing DMDGs have also been put forward. The so‑called ‘mini-Bullet Cluster scenario' proposes that DMDGs arise from a high-speed collision between two gas‑rich dwarf galaxies \citep{silk_ultra-diffuse_2019,shin_dark_2020,lee_dark_2021,van_dokkum_trail_2022,otaki_frequency_2023}. In this framework, the baryonic gas undergoes shock compression, producing overpressurised clouds that trigger intense star formation, while the collisionless DM haloes traverse the collision site and decouple from the gas, leaving behind a remnant that is deficient in DM. The ‘tidal dwarf galaxy scenario’, suggests that DMDGs originate in gaseous, DM‑poor tidal tails produced during mergers of gas‑rich galaxies \citep[e.g.,][]{zwicky_multiple_1956,duc_formation_2000,bournaud_missing_2007,lelli_gas_2015,ploeckinger_tidal_2018}. It should be stressed that these scenarios require rather finely tuned circumstances, such as specific internal properties and merger orbits of the progenitor systems \citep[e.g.,][]{Ogiya2022b}, to reproduce the principal observed characteristics of DF2 and DF4, including the tidal features detected in their outskirts.

Independent of how DF2 and DF4 formed, DMDGs that originate from extreme tidal stripping arise naturally within the standard hierarchical framework of structure formation. Stellar streams and stellar haloes are evidence that tidal effects can penetrate the stellar bodies of satellite galaxies \citep[e.g.,][]{quinn_formation_1984,munoz_modeling_2008}. Consequently, there must exist a class of DMDGs whose outer DM haloes have been stripped off. This paper studies the formation of such systems, with a particular focus on how tidal forces give rise to break radii in the stellar distributions of DMDGs. Furthermore, we argue that break radii observed in the outskirts of typical satellite galaxies are a telltale sign of past tidal encounters and can be employed to identify such systems using photometric data alone. Throughout this paper, when comparing our simulation results to observations, we use DF2 as a benchmark example of a DMDG, even though we acknowledge that there are alternative formation scenarios for DF2, as discussed above.

The paper is structured as follows: In \autoref{sec:model}, we outline the methodology of our $N$-body simulations, detailing the galaxy model, host potential, orbital configuration, and the parameter space explored. \autoref{sec:mock_obs} explains our methods for identifying the break radius and creating mock observations. In \autoref{sec:results}, we present the results from our simulations, starting with a summary of the general evolution of the simulated galaxies and then focusing on the formation, development, and stability of the break radius. We further analyse the break radius by studying its connection to the tidal radius and effective radius in \autoref{sec:rb_model}. Finally, in \autoref{sec:summary}, we summarize our findings and discuss the implications.


\section{Simulation model}
\label{sec:model}

Our simulation configuration utilizes the approach outlined in \cite{ogiya_tidal_2022}. We simulate a satellite galaxy (potential progenitor of DF2) as an $N$-body system that is captured by the host galaxy, NGC\,1052, which is represented using an analytical potential, at accretion redshift $z_{\rm acc}=1.5$. It is assumed that star formation in the satellite galaxy has been completely halted since that time. The corresponding lookback time is approximately equal to the age of stars in DF2, $\sim$9 Gyr \citep{van_dokkum_enigmatic_2018, fensch_ultra-diffuse_2019}. At infall, the satellite galaxy possesses a stellar mass of $2 \times 10^8\ \msun$, and is surrounded by a DM halo with a mass of $6 \times 10^{10}\ \msun$\footnote{The virial overdensity of 200 is adopted throughout the paper, i.e., $M\sub{200} = 6 \times 10^{10}\,\msun$.}, in agreement with the stellar-to-halo mass relationship \citep{behroozi_universemachine_2019}.

\subsection{\texorpdfstring{$N$}{N}-body satellite galaxy model}
\label{sec:N-body galaxy model} 

The $N$-body satellite galaxy model consists of two components: a stellar spheroid and a spherical DM halo. The fiducial structural parameters for both components are chosen to be consistent with observational constraints, while ensuring that the system represents a typical galaxy in terms of its configuration prior to infall.

\subsubsection{Stellar spheroid}

For modelling the stellar spheroid, we use the deprojected form of a S\'{e}rsic profile \citep{ciotti_analytical_1999},
\begin{equation}
    \rho_{*}=\rho_0  \left( \frac{r}{R\sub{e}} \right)^{-p\sub{n}}\exp \left[ -b\sub{n} \left( \frac{r}{R\sub{e}} \right)^{1/n} \right],
    \label{eq:rho_s}
\end{equation}
where $r$ is the distance from the centre of the satellite galaxy, $R\sub{e}$ is the effective radius, $n$ is the S\'{e}rsic index. The parameters depending on $n$ are given as follows \citep{lima_neto_specific_1999},
\begin{equation}
    b\sub{n} = 2n - \frac{1}{3}+\frac{4}{405n}+\frac{46}{25515n^2},
    \label{eq:bn}
\end{equation}
\begin{equation}
    p\sub{n} = 1-\frac{0.6097}{n}+\frac{0.05463}{n^2}.
    \label{eq:pn}
\end{equation}


\subsubsection{DM halo}

For the tidal scenario to be successful in reproducing the extreme DM deficiency observed in DF2 and DF4, it has often been proposed that the progenitor satellite galaxy possesses a cored DM density profile, in agreement with observations of nearby dwarf galaxies \citep{Burkert1995,Swaters2003,Oh2015}, rather than the cuspy DM density profile that generally emerges in DM-only $\Lambda$CDM cosmological simulations\citep{navarro_universal_1997,springel_aquarius_2008,Ishiyama2021}. This condition is supported by numerical simulations showing that dynamical heating of DM particles, driven by fluctuations in the gravitational potential of dwarf galaxies induced by stochastic supernova feedback, reduces the central DM density \citep[e.g.,][]{mashchenko_removal_2006,pontzen_how_2012,di_cintio_dependence_2014, ogiya_corecusp_2014,onorbe_forged_2015,read_dark_2016}, as well as by predictions from alternative DM models, such as self-interacting DM \citep{Vogelsberger2012,Rocha2013,Jiang2023}\footnote{Although DM haloes with cored density profiles are more susceptible to tidal stripping than their cuspy counterparts \citep[][]{Penarrubia.etal.10,ogiya_tidal_2022}, the presence of a core is not a strict requirement for forming DMDGs through tidal stripping. Satellite galaxies residing in cuspy haloes can also end up with an extremely low DM content via tidal stripping, as long as the halo concentration is sufficiently low \citep{amorisco_giant_2019} or the satellite was accreted at an early epoch \citep{Ogiya2021}.}.

We incorporate the alteration of the DM density profile by using the generalized NFW profile presented in \cite{read_dark_2016}. Before the transformation takes place, the mass profile adheres to the NFW model $M_{\rm NFW}(r)$, characterized by the virial mass, $M\sub{200} = 6 \times 10^{10}\,\msun$, and the halo concentration, $c=6.6$, as suggested by the concentration-mass-redshift relation \citep{ludlow_massconcentrationredshift_2016}. Post-transformation, the DM mass becomes
\begin{equation}
    M\sub{DM}(r)=f(r)^{m} M\sub{NFW}(r).
\end{equation}
The transformation function is defined as 
\begin{equation}
    f(r) = \tanh{(r/r\sub{c})},
\end{equation}
which illustrates the alteration in the DM density distribution caused by supernova feedback. The parameter $m$, which varies from 0 to 1, regulates the transformation intensity, transitioning from a cuspy NFW profile to a completely cored profile. The variable $r\sub{c}$ refers to the core radius that determines the extent of the constant density core located at the centre of the DM halo.

\subsection{Host potential}
\label{ssec:host}
We model the host galaxy, NGC\,1052, utilizing an analytical, time-dependent potential conforming to the NFW profile. This potential is completely defined by two parameters: the virial mass $M_{200,\mathrm{host}}$ and the concentration $c_{\mathrm{host}}$. The virial mass $M_{200,\mathrm{host}}$ changes over time as per the DM mass assembly history detailed in \cite{correa_accretion_2015}, while the concentration $c_{\mathrm{host}}$ adheres to the concentration-mass-redshift relation of \cite{ludlow_massconcentrationredshift_2016}. The host halo grows in mass from $\sim 3.5 \times 10^{12} \, \msun$ at redshift $z=1.5$, with $M_{200,\mathrm{host}} \propto z^{-0.7}$ until $z \approx 0.7$. At this point, it attains $M_{200,\mathrm{host}} \sim 7 \times 10^{12} \, \msun$, and subsequently, it continues to grow more slowly, eventually achieving $1.1 \times 10^{13} \, \msun$ at $z=0$. The concentration of the host halo increases almost linearly from $c = 4.8$ at $z = 1.5$ to $c = 6.8$ at $z = 0$.

\subsection{Satellite orbit}
\label{ssec:sat_orbit}
In a host potential with spherical symmetry, the orbit of a satellite galaxy is comprehensively described by two dimensionless parameters. The first parameter is expressed as,
\begin{equation}
    x\sub{c} \equiv r\sub{c}(E)/r\sub{200,host}(z\sub{acc}),
\end{equation}
where $r\sub{c}(E)$ represents the radius of a circular orbit corresponding to the orbital energy $E$, and $r_{200,\mathrm{host}}(z_{\mathrm{acc}})$ is the virial radius of the host halo at the time of accretion, which corresponds to the starting point of the simulations. The second parameter is defined as
\begin{equation}
    \eta \equiv L/L\sub{c}(E),
\end{equation}
with $L$ being the orbital angular momentum of the satellite, and $L\sub{c}(E)$ the angular momentum of a circular orbit with the same orbital energy, $E$. An orbit with a smaller $x\sub{c}$ ($\eta$) implies a more tightly bound (eccentric) orbit. In our standard setup, we use $x\sub{c} = 1$ and $\eta = 0.3$, leading to a pericentric distance of $\sim 20$\,kpc and an apocentric distance of $\sim 300$\,kpc. The orbital period is about 2.2 Gyr.

\subsection{Numerical setup}
\label{ssec:num_setup}
$N$-body simulations are performed using \texttt{GIZMO} \citep{hopkins_new_2015}\footnote{\url{https://bitbucket.org/phopkins/gizmo-public/src/master/}} within a host-centric frame in which the origin of the coordinate system corresponds to the centre of the host potential. We use a Plummer force softening parameter \citep{Plummer_problem_nodate} of $\epsilon = 14$\,pc and an opening angle of $\theta = 0.7$ in the tree-based gravitational force computation \citep{Barnes1986}.

The initial position and velocity of DM and stellar particles are first defined in the satellite-centric frame (origin corresponds to the satellite centre). The initial particle positions are sampled according to their respective (DM or stellar) density profiles using the rejection-acceptance technique. Their initial velocities are drawn from the isotropic phase-space distribution function obtained from the Eddington inversion \citep{eddington_distribution_1916}, based on the total gravitational potential of the stellar spheroid and the DM halo. Then, the centre of the satellite galaxy model is relocated to the apocentre of the orbit with its velocity defined by the pair of orbital parameters as outlined in \autoref{ssec:sat_orbit}.

For simulating the DM halo and stellar spheroid, we employ $1.5 \times 10^6$ particles for DM and $5 \times 10^5$ particles for stars, achieving mass resolutions of $4 \times 10^4\, \msun$ and $4 \times 10^2\, \msun$ respectively. As we show in \appref{app:convergence}, using a resolution study in which we vary the mass of the stellar particles from $4 \times 10^4 \, \msun$ to $40 \, \msun$, our results are properly converged and not affected by potential spurious energy transfer between different particle species \citep[e.g.,][]{ludlow_spurious_2023}.

Based on the numerical criteria proposed by \citet{van_den_bosch_dark_2018} to prevent artificial disruption of substructures, our simulations are permitted to follow the tidal evolution of a satellite until it has lost approximately $99.9 \, \%$ of its initial mass. While these criteria were originally calibrated for cuspy haloes rather than the cored haloes predominantly considered in this work, the satellite galaxy in our simulations generally retains more than $1 \, \%$ of its initial mass, which is comfortably above the critical threshold. This substantial safety margin guarantees that all of our simulations are not affected by spurious numerical disruption.

We utilize the technique outlined by \cite{van_den_bosch_disruption_2018} to track the changes in orbit and mass of the satellite galaxy. This method involves iteratively determining the bound mass of the satellite galaxy, along with the position and velocity of its centre, using the phase-space information of its particles.

\subsection{Parameter space}

In order to study the tidal evolution of satellite galaxies, we performed a series of simulations with different orbital and structural parameters. \autoref{tab:parameter_values} provides a comprehensive summary of the parameters used in our models. Values in bold indicate the baseline values used for our fiducial simulation. Non-bold values indicate values used in simulations in which we varied only that parameter, while keeping all other parameters fixed at their fiducial values. Simulations other than the fiducial one are identified by the parameter that was varied and its non-fiducial value used. For example, ‘Re1750' indicates a simulation in which the effective radius of the stellar component, $R\sub{e,i}$, was set equal to $1750\,\mathrm{pc}$,\ while all other parameters are set to the fiducial values indicated in bold-face in \autoref{tab:parameter_values}. Because dwarf galaxies generally have S\'ersic indices $n < 1$ \citep{carlsten_structures_2021, ordenes-briceno_next_2018}, and values above unity are usually found only in more massive systems, we confine the S\'ersic index in our parameter space to $n < 1$. Nevertheless, exploring a wider range of parameters represents a promising avenue for future research.

We verify that our satellite models remain stable in isolation for at least 10\,Gyr. Both the stellar component and the DM halo preserve their initial structures, and the velocity distributions stay isotropic. The only noticeable evolution is a minor reduction in the central density caused by artificial two-body relaxation, leading to at most a 5 percent increase in the effective radius for all considered parameter sets, which is entirely negligible when compared to the impact of tidal interactions.

\begin{table}
    \centering
    \caption{Parameter values used in the simulations. Fiducial values are shown in bold. From top to bottom, the parameters indicate: initial stellar mass ($M_\star$), initial effective radius ($R\sub{e,i}$), S\'{e}rsic index ($n$), DM inner slope parameter ($m$), DM core radius ($r\sub{c}$), orbital energy parameter ($x\sub{c}$), and orbital angular momentum parameter ($\eta$).}
    \label{tab:parameter_values}
    \begin{tabular}{|l|c|c|c|}
        \hline
        \textbf{Parameter} & \multicolumn{3}{c|}{\textbf{Value}}\\
        \hline
        $M_\star\ (\msun)$ & ${\bf 2 \times 10^8}$ & \multicolumn{2}{c|}{$3 \times 10^8$} \\
        $R\sub{e,i}$ (pc) & {\bf 1250} & 750 & 1750 \\
        $n$ & {\bf 0.64} & 0.5 & 0.8 \\
        $m$ & \multicolumn{2}{c|}{{\bf 1 (cored)}} & 0 (cuspy) \\
        $r\sub{c}$ (pc) & {\bf 2190} & 1560 & 2800 \\
        $x\sub{c}$ & {\bf 1.0} & 0.8 & 1.2 \\
        $\eta$ & {\bf 0.3} & 0.2 & 0.4 \\
        \hline
    \end{tabular}
\end{table}

\section{Mock observations}
\label{sec:mock_obs}
This study primarily examines the stellar distribution in the outskirts of tidally formed DMDGs. When a satellite galaxy encounters the external tidal field of its host galaxy, its stellar body can undergo tidal heating during pericentric passages, resulting in structural alterations. This phenomenon becomes especially significant in the later stages of tidal evolution when the DM halo has been substantially stripped away. As stars move outward from the inner regions of the satellite galaxy, they can increase the surface brightness in the outskirts, giving rise to observable signatures of tidal interactions. Aiming for better comparisons of observations and simulations, we produce mock observations of the simulated satellite galaxy. The analysis includes only stellar particles.

\subsection{2D projected image and surface brightness profile}

\begin{figure}
    \centering
    \includegraphics[width=\linewidth]{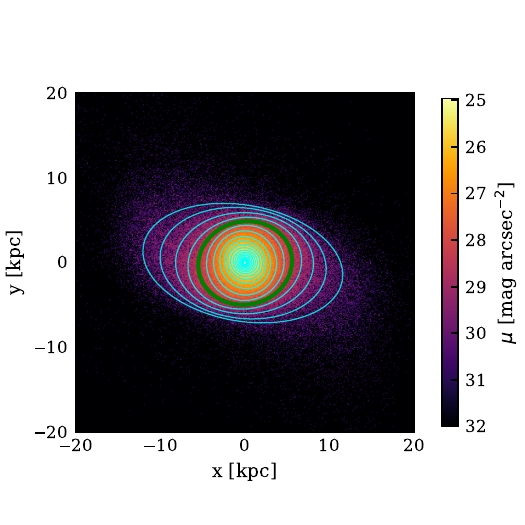}
    \caption{
        Stellar surface brightness map obtained from a mock observation. The snapshot at $t=-2.3\,\mathrm{Gyr}$ from the fiducial simulation is analysed. Pixels fainter than 32 $\text{mag arcsec}^{-2}$ are masked to mimic the detection threshold of observational instruments, such as the Dragonfly Telephoto Array. Cyan lines represent the fitted isophotal contours, while the thick green line marks the isophote corresponding to the break radius.}
    \label{fig:mockob}
\end{figure}

First, the simulated satellite galaxy is projected onto a two-dimensional plane, assuming that it is located at the distance of 22.1\,Mpc measured for DF2 \citep{shen_tip_2021}. The projection plane is then divided into grid cells with a pixel size of $100\,\text{pc} \times 100\,\text{pc}$. Next, we convert the stellar mass in each pixel into a surface brightness by adopting a mass-to-light ratio of $M/L = 2(M/L)_\odot$ \citep{van_dokkum_galaxy_2018}. We then carry out isophotal fitting with \texttt{Photutils}\footnote{\url{https://photutils.readthedocs.io/en/stable/}} \citep{bradley_astropyphotutils_2020}, the Python implementation of the isophote fitting algorithm originally introduced by \cite{jedrzejewski_ccd_1987}. \autoref{fig:mockob} presents a representative mock observation at $t=-2.3\,\mathrm{Gyr}$, where the fitted isophotal contours are indicated by cyan lines. Here, and throughout, $t=0$ corresponds to the present. By this epoch, a distinct stellar stream has developed due to tidal interactions with the host potential, while most of the stars remain gravitationally bound and concentrated near the centre of the satellite galaxy. We adopt an observational detection limit of $32 \, \text{mag arcsec}^{-2}$, following the Dragonfly Telephoto Array \citep{abraham_ultra_2014}, and mask out pixels with fainter surface brightness. For the subsequent analysis, we focus on a $20\,\text{kpc} \times 20\,\text{kpc}$ square region, centred on the centre of the satellite galaxy, containing nearly all the stars in the simulation. The observable stellar mass is then defined as the sum of the stellar mass in all pixels that remain above the detection threshold.

\begin{figure*}
    \centering
    \includegraphics[width=\textwidth]{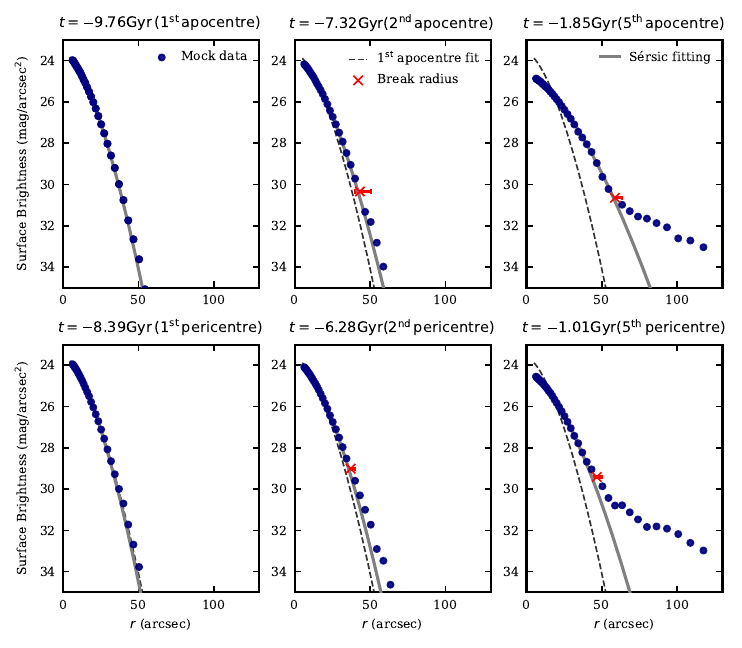}
    \caption{
    Surface brightness profile of the simulated satellite galaxy. Snapshots are provided for the first two and last apocentres (upper panels) and pericentres (lower panels) from the fiducial simulation run. The look-back time is indicated at the top of each panel. For each snapshot, the mock observation analysis produces 100 surface brightness profiles along with 100 associated break radii. The blue points represent the surface brightness for the particular orientation that yields the median break radius. Grey solid line shows the S\'{e}rsic fit to the blue points within 1.5 times the effective radius. Black dashed curve represents the initial S\'{e}rsic profile. The identified break radius, marked with a red cross, is defined as the point where the surface brightness exceeds the S\'{e}rsic fit by at least 0.2 $\text{mag arcsec}^{-2}$. The error bar illustrates the 15–85th percentile range obtained from 100 random projections.
    \label{fig:sfit}
    }
\end{figure*}

From the projected stellar distribution, we derive the surface brightness profile by fitting elliptical isophotes with the \texttt{Photutils} package. The semi-major axes of each successive isophote increase by a factor of 1.1, beginning with the innermost isophote, which has a semi-major axis of 5 arcsec. For each isophote, we compute the mean surface brightness along the corresponding ellipse, thus producing a one-dimensional surface brightness profile that inherently incorporates the ellipticity and position angle of the stellar distribution. The effective radius is determined as the semi-major axis of the ellipse that contains half of the total observable luminosity. Then, we fit the surface brightness profile using the S\'{e}rsic model from 5 arcsec from the centre to 1.5 times the determined effective radius, as described in \cite{keim_tidal_2022}. This analysis is performed for each snapshot from 100 random orientations. \autoref{fig:sfit} illustrates the surface brightness profile of the satellite galaxy model in the fiducial simulation. Blue points in each panel represent snapshots at various orbital phases. Initially, when the galaxy remains largely unaffected by the tidal force of the host, its stellar distribution is accurately represented by a single S\'{e}rsic profile (grey).

\subsection{Break radius} 
Stars that have moved from the inner part of the satellite galaxy due to tidal interactions contribute to a rise in stellar density in the outskirts, causing the surface brightness to exceed the predictions of the inner S\'{e}rsic profile. The point at which this extra brightness becomes visible has been studied and is referred to as the break radius \citep[e.g.,][]{johnston_interpreting_2002, penarrubia_tidal_2008, lokas_tidal_2013, Tang2020, keim_tidal_2022}. The definition of the break radius varies among different studies. According to \cite{lokas_tidal_2013}, when tidal forces are considered, the stellar density profile typically transits from an inner slope of $\rho \propto R^{-6}$ to an outer slope of $\rho \propto R^{-2}$, where $R$ represents the two-dimensional projected distance from the centre of the galaxy. They define the break radius as the point where the logarithmic slope intersects with -4. In contrast, \cite{johnston_interpreting_2002} defines the break radius as the position where the change in the local logarithmic slope of the surface brightness profile between two successive data points exceeds 0.2.

In order to align with observational studies, we employ the definition of the break radius as described by \cite{keim_tidal_2022}. It is defined as the semi-major axis of the innermost isophote at which the surface brightness exceeds the S\'{e}rsic fit by at least 0.2 $\text{mag arcsec}^{-2}$.\footnote{For galaxies with higher S\'{e}rsic indices, determining the break radius from the excess stellar density in the outskirts can be technically more difficult because of their inherent extended envelopes. In such cases, more specialised definitions may be more appropriate.} During the later stages of the simulation (middle and right column panels of \autoref{fig:sfit}), the impact of the host potential becomes evident: the inner stellar density decreases, suggesting that the galaxy is being stretched and expanded by the tidal force of the host. Stars moving outward contribute to the increased surface brightness in the outskirts, causing the profile to deviate from the S\'{e}rsic fit that describes the inner stellar distribution.\footnote{Throughout the fiducial simulation, the S\'{e}rsic index keeps its initial value, $n=0.64$.} The red crosses and error bars in \autoref{fig:sfit} represent the median and the 15th–85th percentile interval of the break radius derived from mock observations generated using 100 random orientation angles. As illustrated in \autoref{fig:mockob}, the outermost isophotes associated with the tidal arms are strongly elongated. In contrast, the isophote that defines the break radius (green ellipse in \autoref{fig:mockob}) remains nearly circular for most of the evolutionary history (see \appref{app:ellipticity}), leading to only minor differences in the break radius between mock observations from various viewing angles.

\section{Results}
\label{sec:results}

\subsection{Global characteristics of the satellite galaxy model}
\label{ssec:glob_char}

\begin{figure*}
    \centering
    \includegraphics[width=\textwidth]{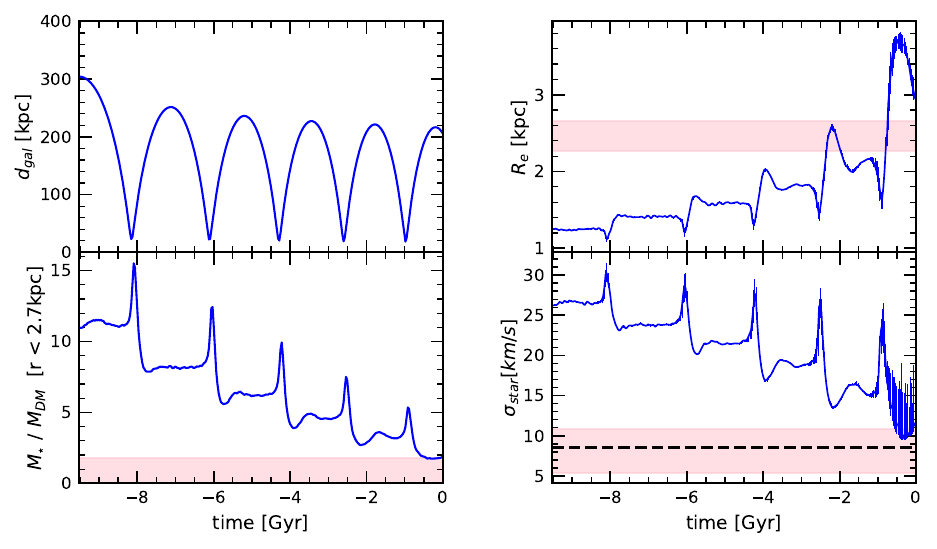}
    \caption{
    Global properties of the satellite galaxy in the fiducial run. Blue line represents the median value obtained from mock observations using 100 random orientations, and the error bars show the 15th to 85th percentile range. 
    \textit{Upper left}: distance between the satellite and the centre of the host halo. The satellite orbit gradually shrinks due to the mass growth of the host.
    \textit{Lower left}: DM-to-stellar mass ratio at a distance of 2.7\,kpc from the centre of the satellite galaxy. The ratio decreases over time as tidal stripping transforms the system from a typical dwarf galaxy to a DMDG. The observational inference is shown by the pink shaded area \citep{danieli_still_2019}.
    \textit{Upper right}: evolution of the effective radius over time. The upper and lower boundaries of the pink shaded region indicate the effective radius of DF2, assuming a distance of 22.1\,Mpc \citep{shen_tip_2021} and 18.9\,Mpc \citep{cohen_dragonfly_2018}, respectively.
    \textit{Lower right}: evolution of the line-of-sight stellar velocity dispersion within the effective radius. The black dashed line and the pink shaded area represent the observed stellar velocity dispersion of DF2, along with its uncertainty \citep{danieli_still_2019}.
    }
    \label{fig:ov}
\end{figure*}

Before examining the break radius of the simulated satellite galaxy in detail, we first analyse how the global properties of the simulated galaxy from the fiducial run evolve in \autoref{fig:ov}. The upper left panel shows the distance between the satellite galaxy and the centre of the host halo. As the simulation progresses, the orbit slowly contracts due to the increasing mass of the host potential \citep{Ogiya2021}. In the lower left panel, the DM-to-stellar mass ratio is depicted, measured at 2.7\,kpc from the centre of the satellite galaxy, equivalent to the three-dimensional half-light radius of DF2 obtained from observations \citep{danieli_still_2019}. Initially around ten, indicating DM dominance before infall, this ratio decreases at each pericentric passage due to tidal stripping, and approaches the observed inference, highlighted as the pink shaded region. The upper and lower right panels display the evolution of the effective radius of the satellite galaxy, and the line-of-sight stellar velocity dispersion measured within the effective radius, respectively. The observed values and their uncertainties are represented by dashed lines and pink shaded areas. These results demonstrate that the present-day satellite in the fiducial simulation effectively captures the overarching dynamical and structural properties of DF2.

\subsection{Emergence and evolution of the break radius}

\begin{figure}
    \centering
    \includegraphics[width=\linewidth]{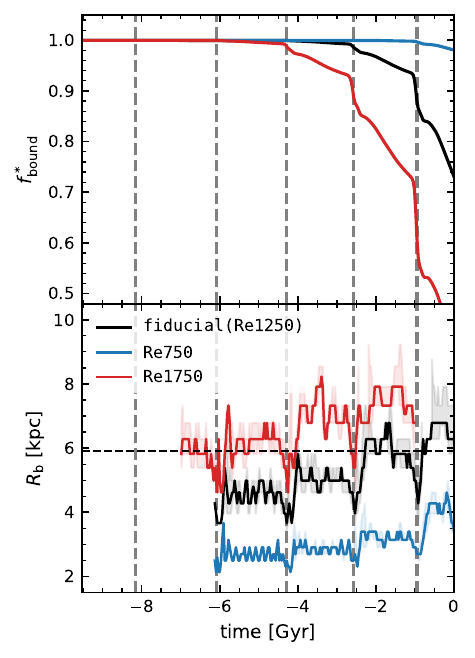}
    \caption{
        Evolution of the stellar component in the satellite galaxy model in simulations varying the initial effective radius of the satellite galaxy model, $R\sub{e,i}$. Results from the fiducial run ($R\sub{e,i} = 1.25$\,kpc) are shown in black, while simulations with $R\sub{e,i} = 0.75$ and 1.75\,kpc are depicted in blue and red, respectively. Vertical dashed lines mark the times of pericentric passages. 
        {\it Upper}: stellar bound mass fraction, $f\sub{bound}^*$. 
        {\it Lower}: break radius, $R\sub{b}$. The lines represent the median from 100 random mock projections, and the shaded areas in matching colours indicate the 15th–85th percentile range. The dashed horizontal line represents the break radius of DF2 \citep{keim_tidal_2022}. 
        In the fiducial run, a distinct break radius emerges after the second pericentric passage, yet the galaxy retains its gravitational hold on all its stars until after the third passage, suggesting that both stripped and loosely bound stars contribute to the stellar distribution beyond the break radius. 
        Simulations with larger $R\sub{e,i}$ consistently show larger $R\sub{b}$ throughout. In the simulations with larger $R\sub{e,i}$, stars begin to be stripped and the break radius appears earlier than in the others. The stellar component is almost disrupted by the end of simulation with $R\sub{e,i}=1.75$\,kpc.
        \label{fig:rbreakRe}
    }
\end{figure}
The black line in \autoref{fig:rbreakRe} shows the time evolution of the stellar bound mass fraction, $f\sub{bound}^*$ (upper), and the break radius, $R\sub{b}$ (lower), of the satellite galaxy in the fiducial simulation. Up to the third pericentric passage ($t \sim -4.3$\,Gyr), all stellar particles remain gravitationally bound, i.e., $f\sub{bound}^* = 1$, whereas DM particles undergo continuous stripping over the course of the simulation. A break radius becomes apparent after the second pericentric passage, before stars are actually stripped from the satellite. Although the tidal force at pericentre is insufficient to directly unbind stars, it dynamically heats the stellar component of the satellite: the stellar velocity dispersion rises and the orbits of stars expand. Stars that migrate outward enhance the stellar density in the outskirts, giving rise to the observed break radius. By the end of the fiducial simulation, the satellite galaxy has become DM deficient (\autoref{fig:ov}) and its break radius comparable to that observed in DF2 (dashed horizontal line in the lower panel). However, the break radius is already clearly visible at earlier stages of the tidal evolution, when the satellite remains DM dominated, indicating that a break radius in the outskirts of satellite galaxies is a typical characteristic of their tidal interactions with their host galaxy. 

Our fiducial simulation successfully replicates key observational characteristics of DF2, as shown in \autoref{ssec:glob_char}. Although other simulations might not replicate these features, they are intended to investigate how changes in model parameters influence the break radius. A summary of the parameters examined in our simulation suite is provided in \autoref{tab:parameter_values}, while the parameters of the host remain constant across all simulations (see \autoref{ssec:host} for further details). The blue and red lines in \autoref{fig:rbreakRe} show the tidal evolution of the stellar component of the satellite galaxy in simulations varying the initial effective radius, $R\sub{e,i}$. We find that the tidal evolution of the stellar body of the satellite galaxy is highly sensitive to $R\sub{e,i}$. Simulations with larger $R\sub{e,i}$ show that the satellite begins to lose its stellar mass earlier than in simulations with smaller $R\sub{e,i}$. Since the initial stellar mass is held fixed across the runs, a larger $R\sub{e,i}$ corresponds to a lower stellar density, which makes the stellar body more susceptible to tidal stripping. Consequently, the outward migration of stars is more pronounced in these cases, yielding larger break radii over the course of the simulations. The influence of other factors, such as orbital parameters and the size of the DM core, on the onset and evolution of the break radius is studied in \appref{app:sim_suite}. 

The lower panel of \autoref{fig:rbreakRe} further illustrates that the break radius shrinks sharply during each pericentric passage and then undergoes a rapid expansion once the satellite moves past pericentre. This cyclical pattern of the break radius closely resembles the behaviour of the effective radius seen in the upper right panel of \autoref{fig:ov}. Near pericentre, the tidal tensor is highly anisotropic and varies quickly. The satellite galaxy becomes elongated along its orbit and compressed in the directions perpendicular to it. These compressive tidal forces raise the central stellar density, leading to a temporary reduction in the apparent size of the system. After the satellite travels away from pericentre, the external compression weakens and the system expands through internal dynamical relaxation. After the expansion phase, the break radius remains approximately constant until the next pericentric passage, exhibiting only mild fluctuations due to the ongoing relaxation in the outskirts of the satellite galaxy. 

\begin{figure*}
    \centering
    \includegraphics[width=\textwidth]{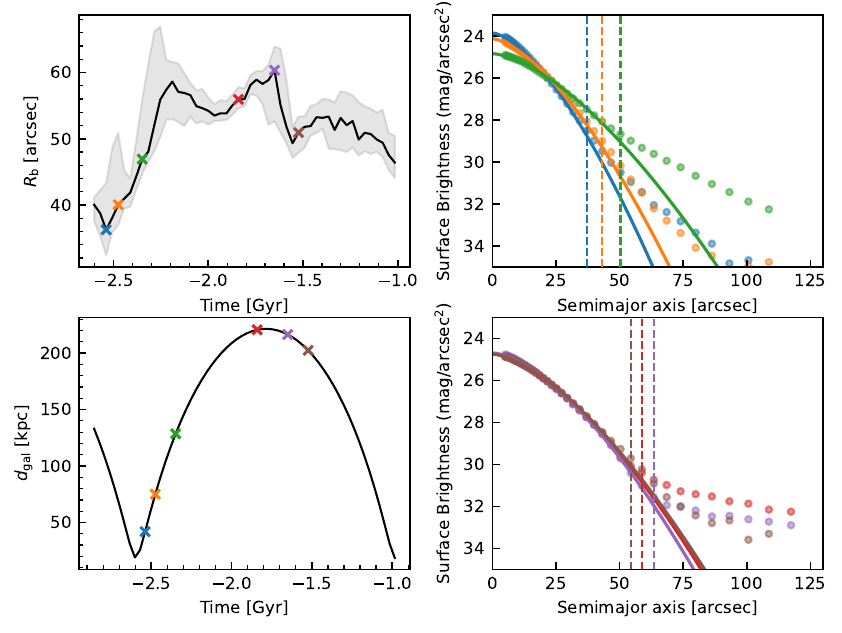}
    \caption{
        A detailed examination of the evolution of the break radius between the fourth and fifth pericentric passages in the fiducial run.  
        \textit{Upper left}: Break radius. The solid line shows the median obtained from 100 random mock projections, and the black shaded region marks the 15th–85th percentile range.  
        \textit{Lower left}: Satellite orbit. Six illustrative snapshots, three near pericentre and three near apocentre, are chosen (crosses) to highlight significant changes in the break radius.   
        \textit{Right}: Surface brightness profiles for these selected snapshots (circles), along with their S\'{e}rsic fits (solid lines) and their corresponding break radii (vertical dashed lines). The colours correspond to those used in the left-hand panels. Profiles with the median break radius values are shown. 
        The rapid and substantial growth of the break radius following the pericentric passage is linked to rapid, strong dynamical relaxation, whereas the smaller variations near apocentre stem from slower and gentler relaxation in the outer regions of the satellite galaxy.
        \label{fig:detail}
    }
\end{figure*}

In \autoref{fig:detail} we study the evolution of the break radius in greater detail, concentrating on its behaviour over a single orbital period between the fourth and fifth pericentric passages. As illustrated in \autoref{fig:rbreakRe}, the break radius increases sharply just after pericentre and then undergoes mild oscillations about a quasi-steady value. To explore the origin of this behaviour, we select six representative snapshots (indicated by coloured crosses in the left panels), three taken close to pericentre and three close to apocentre, that encompass the largest variations in the break radius.

The circles in the right-hand panels of \autoref{fig:detail} show the surface brightness profiles derived from the fiducial simulation, using the same colour scheme as the crosses in the left panels. After the pericentric passage (upper right), the central surface brightness declines because the satellite galaxy expands in response to tidal shock heating. As stars migrate outwards, the surface brightness in the outer regions increases and the break radius grows. This marked transformation in the dynamical structure of the stellar component also leads to significant modifications of the fitted S\'{e}rsic profiles (solid lines): the central surface brightness decreases and the effective radius $R\sub{e}$ increases, implying that the satellite becomes more extended and more diffuse within the break radius.

The lower right panel of \autoref{fig:detail} examines snapshots taken near apocentre, during a phase in which the break radius varies by about $\sim 1$\,kpc ($R\sub{b} \sim 6-7$\,kpc). Although the selected snapshots span a substantial range in break radius, the surface brightness profiles in the inner regions are almost identical, resulting in effectively indistinguishable S\'{e}rsic fits. The discrepancies appear only in the outer regions of the satellite galaxy, where profiles corresponding to smaller break radii display slightly higher surface brightness around the break radius. These modest differences in the outer surface brightness are linked to the slow dynamical relaxation in the outskirts of the galaxy, where the local dynamical timescale is prolonged because of the low density.

\subsection{Stability of the break radius}
To further shed light on the origin and persistence of the break radius, we perform a series of controlled experiments in which the host potential is instantaneously switched off at either pericentre or apocentre. This approach enables us to disentangle the impact of tidal heating from that of ongoing tidal forcing, and to determine whether the presence of a break radius requires a sustained external tidal field.

\begin{figure}
    \centering
    \includegraphics[width=\linewidth]{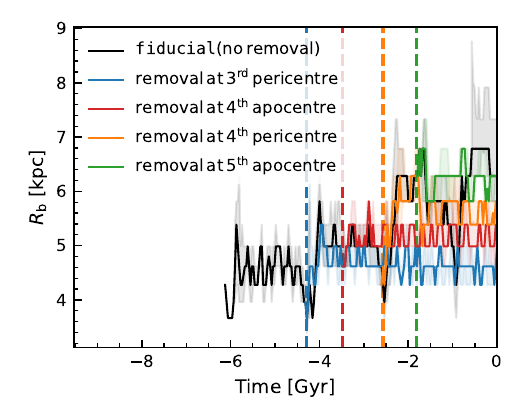}
    \caption{
        Stability test of the break radius. Black line shows the evolution of the break radius in the fiducial run, while in other simulations the host potential is instantaneously removed at the third pericentre (blue), the third apocentre (red), the fourth pericentre (orange), and the fourth apocentre (green). Vertical dashed lines with the corresponding colour indicate the time of the potential removal. The lines and shaded areas indicate the median and the 15th–85th percentile range of break radius measurements from 100 random mock projections in each simulation. When the host potential is removed at pericentre, the amplitude of the break radius variation becomes smaller, suggesting that tidal compression and heating are responsible for the strong changes observed after pericentric passage. The break radius feature remains even when the satellite is no longer subject to the host potential.
    }
    \label{fig:rmp}
\end{figure}
\autoref{fig:rmp} shows how the break radius evolves once the host potential is instantaneously switched off at different orbital phases. When the host potential is removed near pericentre (blue and orange), the subsequent evolution of the break radius qualitatively resembles that in the fiducial run (black): after an initial sharp decline, the break radius expands. Because the satellite is released from the host potential at pericentre, it experiences only about half of the tidal effects, resulting in a smaller variation in the break radius than in the fiducial case. This behaviour confirms that the strong tidal compression and heating around pericentre are responsible for the pronounced fluctuations in the break radius. If the host potential is instead removed at apocentre (red and green), the break radius follows the same evolution as in the fiducial run until the satellite approaches the subsequent pericentre, implying that the host potential has only a minor impact on the satellite near apocentre. In all experiments where the host potential is removed, the break radius settles to an almost constant value once the satellite has dynamically relaxed. We therefore infer that a break radius produced by tidal interactions is imprinted in the stellar distribution of satellite galaxies and remains intact even after they leave the gravitational influence of their host. Consequently, the characteristics of the break radius may provide a promising observational diagnostic for identifying splashback galaxies.

The persistence of the break radius has direct implications for our understanding of galaxy assembly bias and for where we define the dynamical edges of DM haloes. Cosmological simulations indicate that a substantial proportion, possibly as many as 50 percent, of galaxies located just beyond the formal virial radius of galaxy clusters or massive groups are not on their first infall \citep[e.g.,][]{Green.etal.21}. Rather, they are splashback systems that have already passed through pericentre \citep{balogh_origin_2000, gill_evolution_2005}. The prevalence of these interacting systems highlights the importance of the halo boundary definition \citep[][and references therein]{Han2026}. The use of standard spherical-overdensity virial radii can mistakenly classify these orbiting galaxies as isolated field systems \citep{diemer_flybys_2021}. Such galaxies have undergone tidal stripping and environmentally driven quenching of star formation, so their properties differ fundamentally from those of genuine isolated field galaxies and they play a direct role in producing galaxy assembly bias \citep{gao_age_2005}. Identifying individual splashback galaxies in observations is still difficult. Their kinematic properties substantially overlap with those of genuine first-infall systems, so current detections rely on statistical approaches, such as the galaxy cross-correlation function \citep{diemer_dependence_2014, more_splashback_2015}.

The break radius produced by tidal interactions with a host system, which is preserved even after satellite galaxies leave the gravitational potential of the host, can serve as a new diagnostic for identifying splashback galaxies. Satellites located beyond the virial radius of the host that still display a distinct break radius, indicating that they have already undergone intense tidal processing, are strong splashback candidates. In contrast, satellites lacking such a break radius are likely members of the first-infall population. Using this structural diagnostic on observational data from upcoming deep imaging surveys, such as the Chinese Space Station Telescope, the Vera C. Rubin Observatory, and Euclid, could yield direct constraints on the role of environment in driving galaxy assembly bias \citep{zentner_galaxy_2014, zentner_constraints_2019, wang_evidence_2022} and enable a detailed census of the splashback population.

\section{Modelling the evolution of the break radius}
\label{sec:rb_model}

In this section, we seek to determine which physical quantities are most strongly linked to the behaviour of the break radius, and to develop an analytical framework that predicts how the break radius evolves, thereby informing the interpretation of observations.

\subsection{Models in previous studies}
\label{ssec:comp_literatures}

Before examining our simulations in greater depth, we first summarise and evaluate the previously proposed models that predict how the break radius evolves. \cite{lokas_tidal_2013} proposed that the time evolution of the break radius is tightly connected to that of the tidal radius, implying that the emergence of the stellar density excess in the outskirts arises from the response of weakly bound stars to the spatially varying tidal field along the orbit. In their framework, the break radius differentiates an inner region that has largely re-virialised after pericentric passages from an outer region dominated by transient, weakly bound material. However, the tidal radius adopted by \cite{lokas_tidal_2013} assumes that both the satellite and the host can be treated as point masses, an approximation that is clearly over-idealised. We therefore adopt an alternative definition of the tidal radius, explicitly formulated for extended systems, while still testing their qualitative prediction that the break radius correlates with the tidal radius.

Several models for the tidal radius suited to extended systems have been proposed \citep[see][]{van_den_bosch_disruption_2018}. One such model, introduced by \cite{tormen_survival_1998}, is
\begin{equation}
    r\sub{t,1} = R\left[ \frac{m(r\sub{t,1})/M(R)}{2-\frac{\mathrm{d\,ln}\,M}{\mathrm{d\,ln}\,R} \Bigl |_R} \right]^{1/3},
        \label{eq:rt_tormen98}
\end{equation}
where $m(r)$ and $M(R)$ denote the mass profiles of the satellite and host, respectively. Incorporating resonant effects that increase the efficiency of tidal stripping, \cite{klypin_galaxies_1999} obtained an alternative expression for the tidal radius,
\begin{equation}
    r\sub{t,2}=R\left[ \frac{m(r\sub{t,2})}{M(R)} \right]^{1/3}.
        \label{eq:rt_klypin99}
\end{equation}
Following the approach of \cite{klypin_galaxies_1999}, we evaluate $r_{\mathrm{t,1}}$ and $r_{\mathrm{t,2}}$ and adopt the smaller of the two as the tidal radius, $r\sub{K99}$. We also consider the alternative formulation of the tidal radius that includes the centrifugal force as introduced by \citet{King1962},
\begin{equation}
    r\sub{K62} = R \left[ \frac{m(r\sub{K62})/M(R)}{2 + \frac{\Omega^2 R^3}{GM(R)} - \left.\frac{\mathrm{d}\ln M}{\mathrm{d}\ln R}\right|_R} \right]^{1/3} = \left[ \frac{Gm(r\sub{K62})}{\Omega^2 - \frac{\mathrm{d}^2\Phi_{\mathrm{h}}}{\mathrm{d}r^2}} \right]^{1/3},
    \label{eq:rt_K62}
\end{equation}
where $\Omega$ and $\Phi\sub{h}$ denote the angular speed and the gravitational potential of the host, respectively. We further verify that using alternative definitions of the tidal radius \citep[e.g.,][]{read_tidal_2006, read_importance_2006, lokas_tidal_2013, Gajda2016} leads to comparable results.

\cite{penarrubia_tidal_2008} proposed a theoretical framework to estimate $R\sub{P08}$, defined as the radius where the inner, smooth stellar component transitions to the outer profile in which the stellar density exceeds the straightforward extrapolation of the inner distribution. This quantity plays the same qualitative role as the break radius investigated in this work, although the precise operational definition, which is not explicitly stated in their work, may differ. In their picture, stars inside $R\sub{P08}$ have had enough time to dynamically relax following the most recent pericentric passage, whereas the outer regions are populated mainly by weakly bound stars that are still moving outward and have not yet settled into a new equilibrium configuration. Under this assumption, the transition radius is estimated as the distance a particle with velocity $\sigma\sub{0}$ travels after its most recent pericentric passage,
\begin{equation}
    R\sub{P08}(t) = \sigma\sub{0} (t - t\sub{peri}),
        \label{eq:req}
\end{equation}
where $t\sub{peri}$ is the time of the last pericentric passage. For $\sigma\sub{0}$, the central line-of-sight stellar velocity dispersion is adopted as suggested in \cite{penarrubia_tidal_2008}.

\begin{figure}
    \centering
    \includegraphics[width=\linewidth]{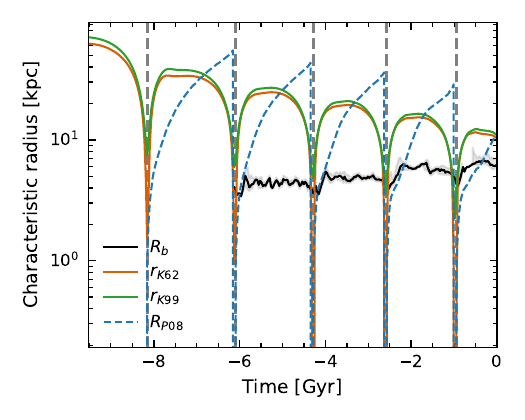}
    \caption{
        Comparison of characteristic radii. Black line and shaded region indicate the median and the 15th–85th percentile range of the break radius, $R\sub{b}$, obtained from 100 mock observations of the fiducial simulation. Orange and green lines show the tidal radius, a quantity that may correlate with the break radius, computed following the prescriptions of \citet{King1962} and \citet{klypin_galaxies_1999}, respectively. Blue dashed line represents the model proposed by \citet{penarrubia_tidal_2008} to describe the evolution of the characteristic scale, which plays a qualitatively similar role to that of the break radius. Vertical dashed lines indicate the moments of pericentric passages. None of these models captures the qualitative behaviour of the break radius.
        }
    \label{fig:Rt}
\end{figure}

\autoref{fig:Rt} compares the break radius obtained from the fiducial simulation (black) with the predictions of the models introduced above. Since the underlying definitions of these quantities differ from that of the break radius, they cannot be compared in a strict quantitative sense. Nonetheless, \autoref{fig:Rt} allows for a qualitative comparison of their behaviour. Although the global evolution of the tidal radius and of the break radius differ, certain correlations become apparent, particularly during pericentric passages. The tidal radius undergoes a sharp decrease at each pericentre and then grows substantially as the satellite travels towards the outskirts of the host, because it is set by the instantaneous balance of forces along the orbit. In contrast, the evolution of the break radius is governed by the combined effects of the intense tidal interaction at pericentre and the subsequent internal relaxation. At every pericentric passage, the break radius shows an immediate decline, mirroring the behaviour of the tidal radius, but the magnitude of this decrease is more modest. After pericentre, the energy injected by the strong tidal encounter causes the break radius to expand quickly, followed by a stage during which it remains roughly constant until the next pericentric approach. The dashed blue line shows that $R\sub{P08}$ predicted by \autoref{eq:req} evolves differently from the break radius\footnote{It should be emphasized that the satellite model adopted by \citet{penarrubia_tidal_2008} (a King stellar spheroid embedded within an NFW halo) differs substantially from those considered here. In their case, the satellite experiences a gradual reduction in both the half-light radius and the break radius after each orbit, which is the opposite behaviour to our results.}. As implied by \autoref{eq:req}, $R\sub{P08}$ drops to zero at every pericentre and then increases monotonically until the next pericentric passage. Note that $R\sub{P08}$ accurately reproduces the break radius immediately after pericentric passages, reinforcing our interpretation that the evolution of the break radius is governed by the internal dynamical relaxation of the satellite galaxy. Based on this analysis, we conclude that no model in the literature captures the evolution of the break radius throughout the entire orbital phase.

\subsection{Comparison of the break radius and the effective radius}
\label{ssec:relation_rbre}

The discussion in \autoref{ssec:comp_literatures} motivates us to construct a new model to forecast how the break radius evolves. We focus on the relationship between the break radius and the effective radius of the satellite galaxy. Unlike the tidal radius, which is determined externally by the gravitational potential of the host, the effective radius characterises the internal structure of the satellite and the spatial distribution of its stars, as with the break radius.

\begin{figure}
    \centering
    \includegraphics[width=\linewidth]{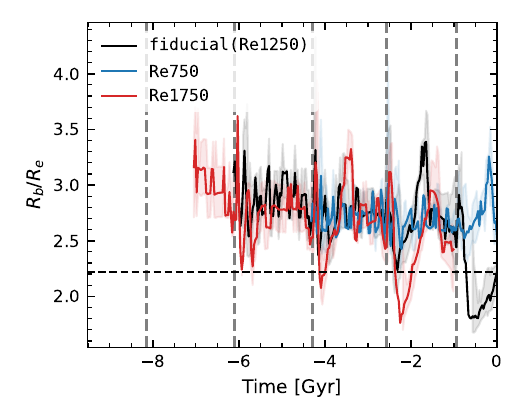}
    \caption{
        Break radius-to-effective radius ratio, $R\sub{b}/R\sub{e}$, as a function of time. Horizontal dashed line is the ratio obtained from observations of DF2, while vertical dashed lines represent the time of pericentric passages. Results from the fiducial simulation with $R\sub{e,i} = 1.25$\,kpc are shown in black, while simulations with $R\sub{e,i} = 0.75$ and 1.75\,kpc are indicated in blue and red, respectively. Solid lines show the median values from 100 mock observations with randomly sampled orientation angles, while the shaded areas indicate the 15th–85th percentile distribution. Despite the substantial variation in $R\sub{b}$, the ratio consistently approaches a value of about 2.7 throughout the simulations, except in regimes where the stellar component undergoes severe tidal stripping. 
    }
    \label{fig:RbRe}
\end{figure}

\autoref{fig:RbRe} presents the time evolution of the ratio between the break radius and the effective radius, $R\sub{b}/R\sub{e}$, for simulations with varying initial effective radii, $R\sub{e,i}$. Although the associated break radii differ substantially, as illustrated in \autoref{fig:rbreakRe}, the ratio $R\sub{b}/R\sub{e}$ stays narrowly constrained between 2.5 and 3 throughout most stages of the simulations. This behaviour indicates that the break radius is primarily set by the dynamical response of the satellite galaxy to tidal heating, which drives both the outward migration of stars and the growth of the effective radius. A more extensive comparison across a broader set of initial configurations is provided in \appref{app:sim_suite}. Despite variations in the initial conditions, $R\sub{b}/R\sub{e}$ tends to settle around a value of $\sim 2.7$. However, it can reach lower values, down to $\sim 2$, when the stellar component of the satellite is heavily disrupted by tidal forces, as occurs, for instance, after the fifth pericentric passage in the fiducial simulation. Recent observational results offer useful reference points for our findings: \cite{keim_tidal_2022} report a break radius of $55^{\prime\prime}$ and an effective radius of $24.8^{\prime\prime}$, yielding $R\sub{b}/R\sub{e} = 2.21$. This value matches the last stage, corresponding to the current epoch, of the fiducial simulation, at which point the satellite exhibits several properties similar to DF2 (\autoref{fig:ov}).

Given that the ratio $R\sub{b}/R\sub{e}$ remains approximately constant, the evolution of the break radius closely follows that of the effective radius of the stellar mass distribution. As originally shown by \citet{Hayashi.etal.03} and \citet{penarrubia_tidal_2008}, the tidal evolution of DM subhaloes follows a well-defined path in structural parameter space. In particular, they showed that the size and internal kinematics of subhaloes follow tight `tidal tracks' that only depend on the properties at infall and the amount of mass lost \citep[e.g.,][]{Green.vdBosch.19, errani_asymptotic_2021, stucker_tidal_2023, Aguirre-Santaella2026}. Building on this framework, \cite{errani_systematics_2018} and \citet{carleton_formation_2019} refined these relations and demonstrated that the stellar and DM components of satellite galaxies evolve along distinct characteristic trajectories, with the stellar tracks being particularly sensitive to the initial degree of spatial segregation between stars and DM. In this work, we employ the tidal track introduced by \citet{errani_systematics_2018} to model the evolution of the effective radius of satellite galaxies experiencing tidal interactions, according to which
\begin{equation}
    R\sub{e} = R\sub{e,i} \frac{(1+x\sub{s})^\alpha x^\beta}{(x+x\sub{s})^\alpha},
        \label{eq:tidal_track_carleton19}
\end{equation}
with $\alpha = -0.23$, $\beta = -0.23$, and $x\sub{s} = 1$. Here, $R\sub{e,i}$ is the initial effective radius, and $x$ quantifies the degree of tidal mass loss,
\begin{equation}
    x \equiv \frac{M\sub{evolved}(r_{\mathrm{max}})}{M\sub{infall}(r_{\mathrm{max}})},
\end{equation}
where $M\sub{infall}(r)$ and $M\sub{evolved}(r)$ denote the dynamical mass profiles at infall and after the tidal interaction, respectively, and $r\sub{max}$ is the radius from the satellite centre at which the circular velocity peaks. Since the mass profile changes throughout the tidal evolution, $r\sub{max}$ itself varies with evolutionary stage.

\begin{figure}
    \centering
    \includegraphics[width=\linewidth]{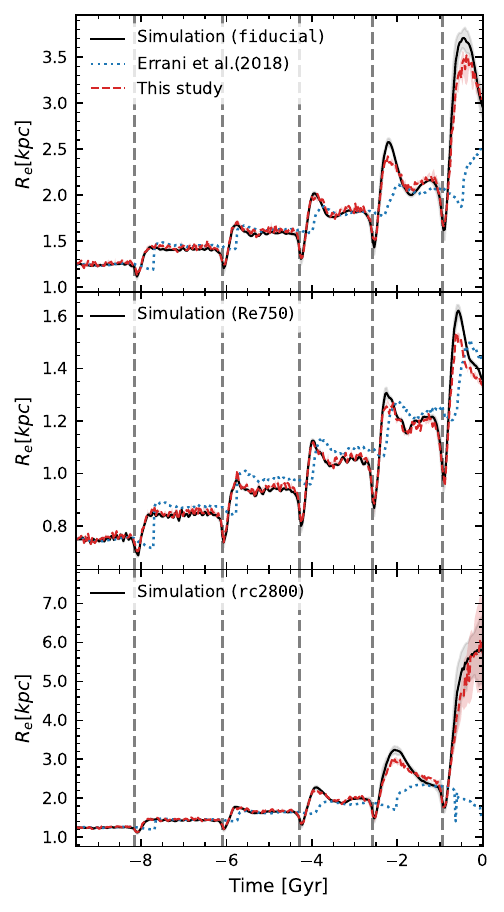}
    \caption{
        Comparison between the evolution of the effective radius obtained from simulations and that predicted by tidal track models. Each panel analyses a model with a different internal structure for the satellite galaxy. Black solid line and black shaded band correspond to the median and the 15th–85th percentile range derived from 100 mock observation analyses of the simulations. Blue dotted line shows the predictions from the tidal track model of \citet{errani_systematics_2018}. Red dashed line and red shaded region represent the median and 15th–85th percentile interval of the predictions from our new tidal track. Vertical dashed lines mark the times of pericentric passages. Our updated tidal track model accurately reproduces the results obtained from the simulations. 
    }
    \label{fig:CarletonRe}
\end{figure}

\autoref{fig:CarletonRe} compares the prediction from \autoref{eq:tidal_track_carleton19} (blue dotted) with the effective radius obtained from our simulations (black solid). The tidal track reproduces the simulation outcome well throughout most evolutionary stages, but deviates near pericentric passages and in the final phase of the simulations, when the satellite galaxy departs from dynamical equilibrium. In both the simulations and the tidal track model, the effective radius shows a characteristic pattern of contraction followed by expansion after each pericentric passage. However, the tidal track underestimates the amplitude of these variations and its predicted pattern generally lags behind that seen in the simulations. This discrepancy arises because the tidal track is calibrated at $r_{\mathrm{max}}$, where the local dynamical timescale is longer and the dynamical response is weaker than in the inner galactic regions that determine the effective radius. The tidal track proposed by \cite{errani_systematics_2018} fails to match the $R\sub{e}$ evolution found in our simulations at late times. Although their model is constructed for DM-dominated satellites, the satellite galaxy in our simulations becomes a DMDG after a few orbital periods, which causes the discrepancy.

\begin{figure}
    \centering
    \includegraphics[width=\linewidth]{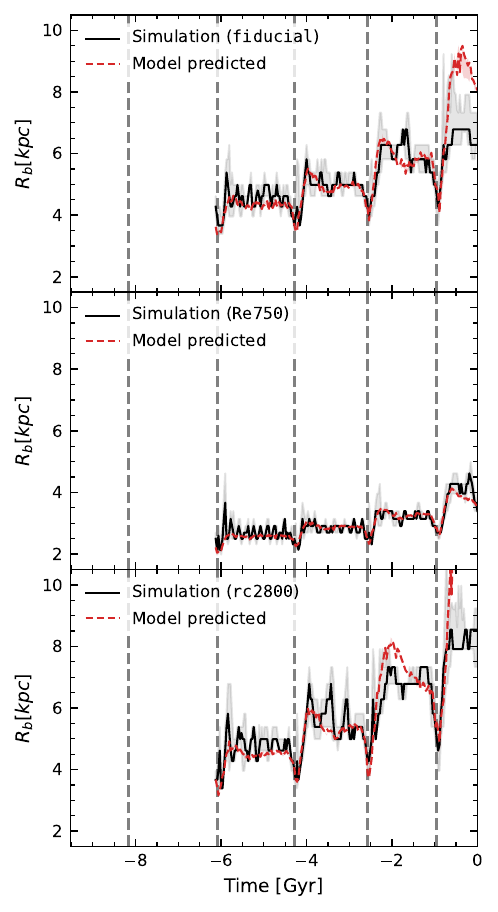}
    \caption{
        Comparison of the evolution of the break radius derived from the simulations to that predicted by our model. The stellar particles in the simulations are analysed from 100 random orientations to get the median (black solid line) and the 15-85 percentile distribution (black band). Red dashed line and the boundaries of the red shaded region represent the break radius predicted by our prescription, using the central surface brightness derived from mock observations taken at the same orientation angles as those used for the black line and the boundaries of the black band. Our model accurately tracks the evolution of the break radius over the course of the tidal evolution of the satellite galaxy, with the exception of the final stage of the simulations, when the stellar component becomes heavily disrupted and $R\sub{b}/R\sub{e}$ departs from the typical value.}
    \label{fig:ModelRb}
\end{figure}

After exploring different choices for $x$ in \autoref{eq:tidal_track_carleton19}, we found 
\begin{equation}
    x \equiv \frac{I\sub{0,evolved}}{I\sub{0,infall}},
\end{equation}
with $\alpha = 0.18$, $\beta = -0.42$, and $x\sub{s}=1$ provides better matches to the simulations (red dashed line in \autoref{fig:CarletonRe}). Here, $I\sub{0,infall}$ and $I\sub{0,evolved}$ denote the central surface brightness at infall and after tidal evolution, respectively. The central surface brightness better captures the rapid dynamical relaxation at $R\sub{e}$, owing to the short dynamical timescale in the galactic centre. In contrast, $r\sub{\max}$ typically lies beyond $R\sub{e}$, where the local dynamical time is longer and therefore less faithfully tracks the rapid evolution of $R\sub{e}$. Consequently, the new tidal track defined by the central surface brightness outperforms those relying on $r\sub{\max}$. As illustrated in \autoref{fig:RbRe}, the break radius to effective radius ratio remains approximately constant at $R\sub{b}/R\sub{e} \sim 2.7$ in all simulations. By multiplying the predicted $R\sub{e}$ by a factor of 2.7, the revised tidal track also successfully matches the evolution of the break radius found in the simulations (\autoref{fig:ModelRb}). However, this prescription tends to overshoot $R\sub{b}$ in cases where the stellar component of the satellite galaxy is heavily disrupted, when the ratio can fall to $R\sub{b}/R\sub{e} \sim 2$. Because the factor of 2.7 is always multiplied to the predicted $R\sub{e}$, the inferred break radius eventually becomes larger than the value measured in the simulations.

A series of studies has carried out an in-depth analysis of the tidal tracks of the stellar components of satellite galaxies \citep[][]{errani_systematics_2018, errani_can_2020, errani_structure_2022}, showing that the characteristics of stellar tidal remnants depend on both the stellar binding energy and the spatial segregation between stars and DM, quantified by
\begin{equation}
    s \equiv R\sub{e,i}/r\sub{max},
        \label{eq:segregation}
\end{equation}
where $R\sub{e,i}$ and $r\sub{max}$ are the initial stellar effective radius and the radius of maximum circular velocity, respectively. In our simulations, the DM core is much smaller than the scale radius of the original NFW profile. As a result, $r\sub{max}$ is barely affected by the modification of the DM density profile in our setup, and thus $s$ is effectively governed by $R\sub{e,i}$. Across our simulation suite, $s$ varies from $\sim 1/20$ to $\sim 1/10$, with $s = 0.085$ in the fiducial run. The parameters in the tidal track for the effective radius derived by \citet{errani_systematics_2018} were calibrated for $s = 1/20$ and $1/10$, and they exhibit only a weak dependence on $s$. This insensitivity over $s = [1/20:1/10]$ may account for why a single tidal track for the effective radius is able to accurately describe all of our simulations.

\section{Summary}
\label{sec:summary}
Through interactions with the host galaxy, stars and dark matter (DM) are stripped from the outskirts of satellite galaxies. At the same time, stars in the central regions are gradually pushed out to larger radii by dynamical heating. This process produces an enhanced surface brightness in the outer parts of the satellite relative to the extrapolation of the inner S\'{e}rsic profile. Breaks of this kind in the surface brightness profile have been identified in both observational data and numerical simulations. The recently identified DM deficient galaxies (DMDGs), DF2 and DF4, are particularly compelling objects, as their key properties, such as their low DM content, large effective radii, and small stellar velocity dispersions, can be naturally explained if they originated through strong tidal stripping. Moreover, observations reveal evidence of tidal interactions in both DF2 and DF4, as indicated by the characteristic behaviour of the break radius in their surface brightness profiles. In this work, we use controlled $N$-body simulations of interacting galaxies to investigate how this break radius forms and evolves over time, focusing on DMDGs as a representative case of satellite galaxies that have been strongly perturbed by tides. Our key results are summarised below.

\begin{itemize}
    
\item Tidal stripping can generate DMDGs if the DM halo of the progenitor satellite exhibits a cored central density profile. Tidal forces initially peel away the weakly bound DM in the outer regions of the satellite, whereas the centrally concentrated stars, embedded in a deeper potential well, are largely unaffected. In the fiducial simulation, no stars become unbound until after the third pericentric passage. Tidal heating injects kinetic energy into the stellar component, producing sharp spikes in the stellar velocity dispersion after each pericentric passage. As a result, once the satellite travels away from pericentre, the stellar distribution expands rapidly, leading to a substantial growth in the effective radius. In contrast, along the calmer parts of the orbit, far from pericentre, the effective radius only varies modestly. 
    
\item Our simulations produce a break radius with sizes comparable to those observed in DF2 and DF4. The break radius changes over time, typically growing as the satellite is gradually heated by tidal interactions with the host galaxy. At each pericentric passage, strong tidal compression causes a temporary reduction in the break radius, which is then followed by a rapid expansion phase as the satellite undergoes dynamical relaxation. During the calmer orbital phases, farther from pericentre, the break radius shows only mild variations, driven by the continued dynamical relaxation in the outskirts of the satellite.
    
\item The evolution of the break radius strongly depends on both the initial effective radius of the progenitor satellite galaxy and the degree of tidal stripping. Having a larger initial effective radius, or when the satellite undergoes more severe tidal mass loss, a larger break radius forms. \appref{app:sim_suite} presents simulation results in which other parameters are varied, such as the satellite orbit relative to the host and its DM density profile.
   
\item The tidal field is not necessary to preserve a break radius, even though the break itself originates from tidal interactions. When the satellite galaxy is released from the host potential at pericentre, the break radius grows, showing the same qualitative behaviour as in simulations where the tidal field remains active. This demonstrates that the post-pericentric growth of the break radius is primarily driven by the internal relaxation of the satellite, rather than by continued tidal forcing. Similarly, when the host potential is removed at apocentre, the break radius remains essentially constant for the remainder of the simulation. This persistence implies that once a break radius is generated by tidal interactions, it remains intact as a fossil imprint long after the satellite escapes the gravitational influence of the host. We therefore propose that the break radius can be used as a powerful morphological diagnostic to identify splashback galaxies beyond the formal virial radius, thereby providing a crucial tool for isolating environmental effects and measuring galaxy assembly bias.
    
\item The break radius, $R\sub{b}$, is linked to the effective radius, $R\sub{e}$, such that their ratio remains nearly constant over the course of the evolution, with $R\sub{b}/R\sub{e} \approx 2.7$. When the stellar component of the satellite undergoes strong tidal disruption, the ratio can decrease to $\sim 2$, which agrees with the observationally inferred value for DF2. A new tidal track, which characterizes the evolution of $R\sub{e}$ in terms of the variation in the central surface brightness of the satellite galaxy, reproduces the simulation outcomes with high accuracy. When this relation is combined with the constant $R\sub{b}/R\sub{e}$ ratio, it provides a predictive framework for the time evolution of the break radius.
\end{itemize}

Distinguishing break-radius features from intrinsic multiple stellar components in galaxies, or from stellar haloes formed in past mergers, may require deep spectroscopic observations to obtain the detailed kinematics of the galaxies. The outer stellar regions of satellite systems, including DF2 and DF4, exhibit notable characteristics that could indicate tidal interactions with their host, including S‑shaped surface-brightness structures, position‑angle twists, and increasing isophotal ellipticity. Comprehensive theoretical investigations of these additional tidal signatures could facilitate this distinction and therefore constitute a promising avenue for future research.

\section*{Acknowledgements}
We thank the anonymous referee for providing insightful comments. ZY and GO were supported by the National Key Research and Development Program of China (No. 2022YFA1602903), the National Natural Science Foundation of China (No. 12373004, W2432003), and the Fundamental Research Fund for Chinese Central Universities (No. NZ2020021, 226-2022-00216). FvdB was supported by the NSF through grant AST 2407063. This study was funded by the Zhejiang University Global Partnership Fund. We acknowledge the cosmology simulation database in the National Basic Science Data Center (NBSDC) and its funds, the NBSDC-DB-10.

\section*{Data Availability} 
The data and code underlying this article will be shared on reasonable request to the corresponding author.

\bibliographystyle{mnras}
\bibliography{ref}



\appendix

\section{Numerical Convergence}
\label{app:convergence}

\begin{figure}
    \centering
    \includegraphics[width=\linewidth]{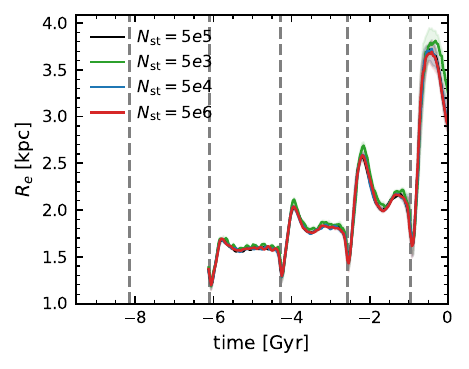}
    \caption{
        Evolution of the effective radius, $R\sub{e}$, obtained from simulations with different numbers of stellar particles, $N\sub{st}$. Black, green, blue, and red curves correspond to simulations with $N\sub{st} = 5 \times 10^5$ (the standard resolution used in this work), $5 \times 10^3$, $5 \times 10^4$, and $5 \times 10^6$, respectively. For each case, the solid line and the shaded band of the same colour represent the median and the 15th–85th percentile interval, respectively, as obtained from 100 mock observation analyses. The values of $R\sub{e}$ in our simulations show good numerical convergence.
        }
    \label{fig:convergence}
\end{figure}

\autoref{fig:convergence} presents a convergence test for the effective radius based on simulations that vary the number of stellar particles, i.e., the stellar mass resolution, while keeping all other numerical parameters fixed as described in \autoref{ssec:num_setup} and adopting the same physical parameters as in the fiducial run. The effective radii measured in these simulations are in good mutual agreement, indicating that our simulation results are numerically converged.

\section{Ellipticity at the break radius}
\label{app:ellipticity}

\begin{figure}
    \centering
    \includegraphics[width=\linewidth]{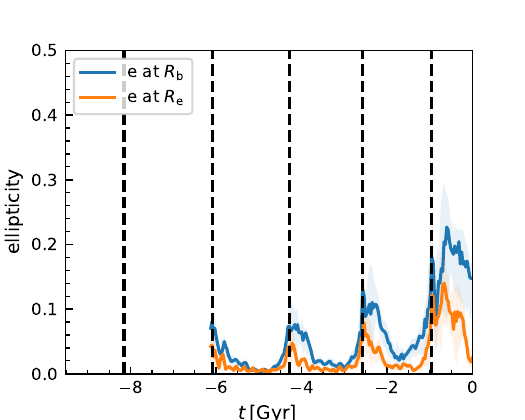}
    \caption{
        Evolution of the ellipticity of the isophotes whose semi-major axis corresponds to the break radius $R\sub{b}$ (blue) and the effective radius $R\sub{e}$ (orange). The solid lines and shaded regions show, respectively, the median and the 15th–85th percentile interval derived from 100 mock observational realisations of the fiducial simulation. Vertical dashed lines mark the times of pericentric passages. The isophotes at $R\sub{b}$ and $R\sub{e}$ remain almost circular throughout most of the evolution, deviating only in the final phase of the simulation, when the stellar component of the satellite is heavily disturbed by tidal forces.
        }
    \label{fig:ell}
\end{figure}

\autoref{fig:ell} shows the evolution of the ellipticity of the isophotes whose semi-major axis defines the break radius (blue) and the effective radius (orange) in the fiducial run. Whenever the satellite approaches a pericentre, the ellipticity rises sharply, driven by a temporary distortion of the stellar body of the satellite caused by the strong tidal field. Following the pericentric passage, the ellipticity declines again. For most of the evolution, the ellipticity at the break radius remains below 0.1, implying only minor differences among mock observations generated from different viewing angles. During the final phase of the simulation, when the satellite galaxy has lost a significant amount of DM mass and has been transformed in a DMDG, the ellipticity exceeds 0.2, producing a larger viewing-angle-induced scatter in the measured break radius (see, e.g., the lower panel of \autoref{fig:rbreakRe}).
Although the elevated ellipticity in this epoch causes considerable scatter in the break radius, the scatter in the ratio $R\sub{b}/R\sub{e}$ does not increase correspondingly, because the ellipticity at $R\sub{e}$ evolves in tandem with that at $R\sub{b}$.

\section{Entire simulation suite}
\label{app:sim_suite}

\begin{figure*}
    \centering
    \includegraphics[width=\textwidth]{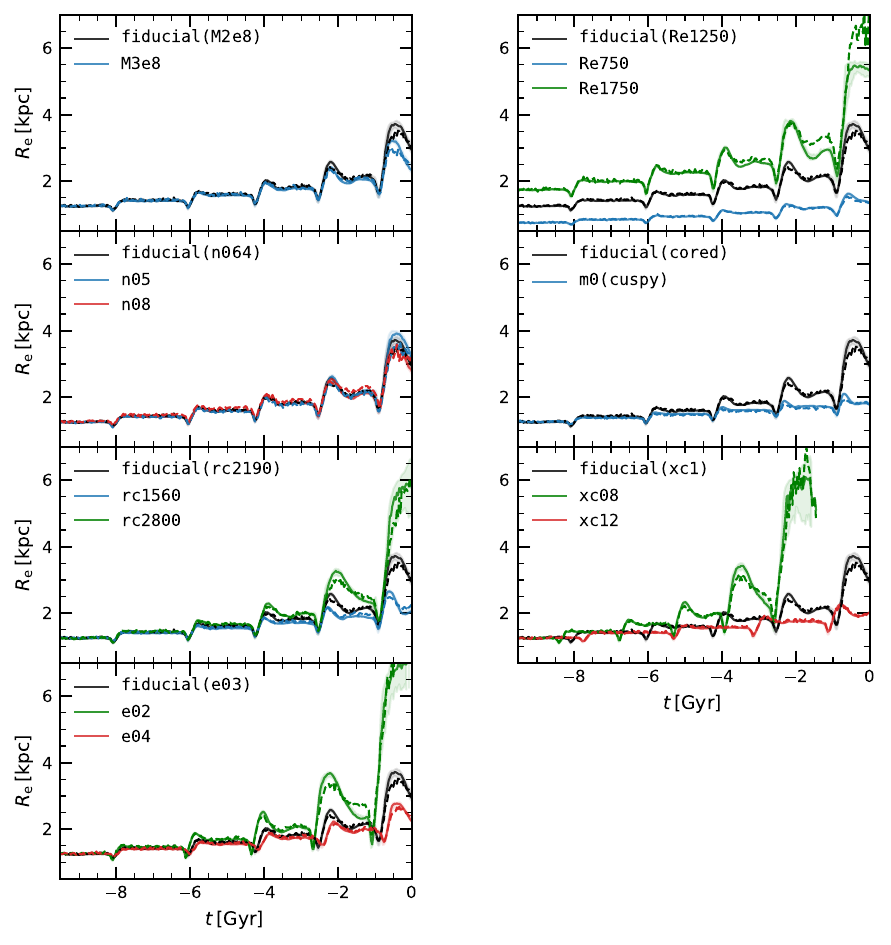}
    \caption{
        Evolution of $R\sub{e}$ in the entire simulation suite. The solid curve shows the median value derived from 100 mock observations taken from random orientations, while the shaded area of the same colour represents the 15th–85th percentile range. Our tidal track (dashed curve), defined via the central surface brightness derived from the mock observation taken at the same orientation angle as that used for the solid line of the same colour, faithfully reproduces the simulation results over a broad range of internal DM halo and stellar configurations in the satellite galaxy, as well as a variety of orbits within the host potential. In the simulations highlighted in green, the satellite galaxy approaches disruption.
    }
    \label{fig:fullRb}
\end{figure*}

\begin{figure*}
    \centering
    \includegraphics[width=\textwidth]{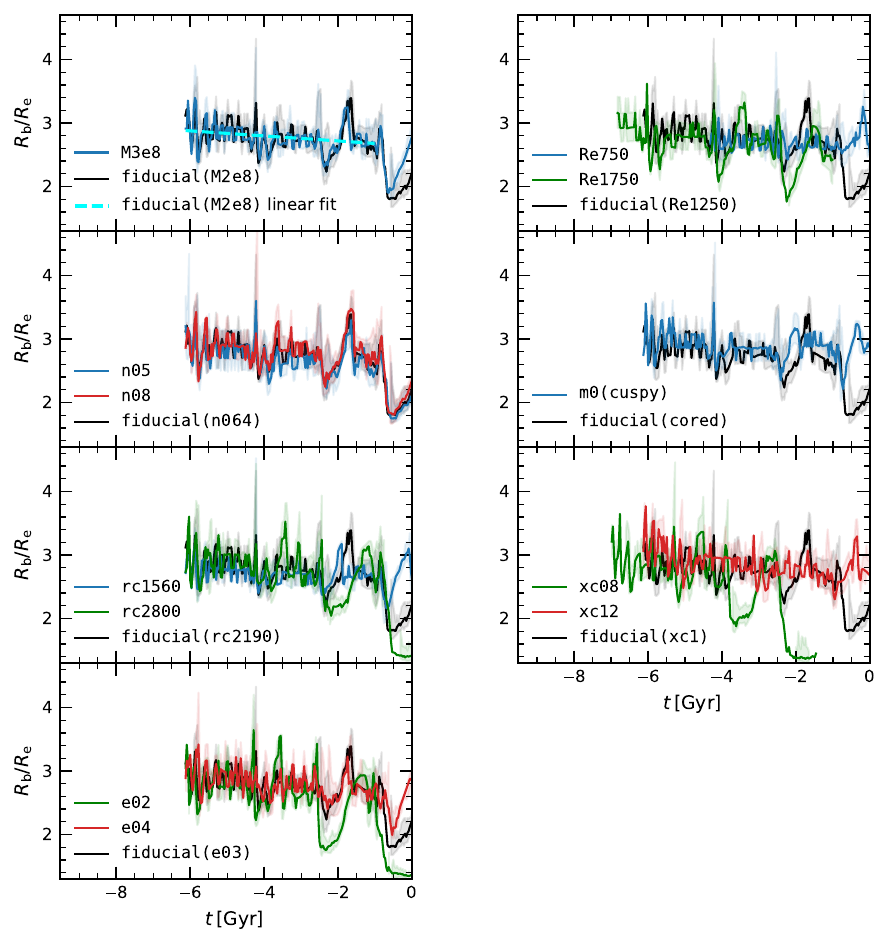}
    \caption{
        Break radius to effective radius ratio, $R\sub{b}/R\sub{e}$, across the simulation suite. $R\sub{b}/R\sub{e}$ remains at $\sim 2.7$, but it exhibits an abrupt decline when the satellite galaxy approaches complete disruption (shown in green). Cyan line in the top-left panel indicates the linear fit.
    }
    \label{fig:fullRbRe}
\end{figure*}

In this Section, we present the results from the full simulation set, where we vary the structural parameters of the DM halo and the stellar component of the satellite galaxy model, as well as the orbital configuration within the host. \autoref{fig:fullRb} shows the evolution of the effective radii across the entire parameter space explored in our simulation suite (solid). Our tidal track based on the central surface brightness of the satellite galaxy (dashed), provides a good match to the simulation results. Furthermore, \autoref{fig:fullRbRe} illustrates that the ratio between the break radius and the effective radius remains generally constant at $R\sub{b} / R\sub{e} \sim 2.7$, consistent with the conclusion in \autoref{ssec:relation_rbre}. Once the satellite galaxy approaches complete disruption (green), this ratio undergoes a sharp decline.

To quantitatively confirm the visually inferred constancy of $R\sub{b} / R\sub{e}$, we perform a linear fit to the evolution of $R\sub{b}/R\sub{e}$ using the NumPy library \citep{harris_array_2020},
\begin{equation}
    R\sub{b}/R\sub{e} = kt + b,
        \label{eq:linear_fit}
\end{equation}
excluding the final, strongly disrupted stage (i.e., the period following the last pericentric passage). The linear fit for the fiducial run is shown as a cyan line in the top-left panel of \autoref{fig:fullRbRe}, and the corresponding fit parameters for the remaining simulations are listed in \autoref{tab:linear_fits}. The resulting fits exhibit extremely shallow slopes (mean $k \approx -0.04\,\mathrm{Gyr}^{-1}$), implying that the structural profile undergoes only minimal secular evolution and has effectively reached a stable configuration, with an average intercept of $b \approx 2.61$, in agreement with the value inferred visually.

\begin{table}
    \centering
    \caption{
    Linear fit parameters describing the evolution of $R\sub{b}/R\sub{e}$. The intercept $b$ corresponds to the asymptotic convergence value for each model variation, and $k$ denotes the slope of the linear relation.}
    \label{tab:linear_fits}
    \begin{tabular}{lcc}
        \toprule
        Model & Intercept ($b$) & Slope ($k$) \\
        \midrule
        $\mathtt{fiducial}$ (Baseline) & $2.638$ & $-0.040$ \\
        \addlinespace
        $\mathtt{n05}$ & $2.507$ & $-0.056$ \\
        $\mathtt{n08}$ & $2.758$ & $-0.021$ \\
        \addlinespace
        $\mathtt{M3e8}$ & $2.619$ & $-0.040$ \\
        \addlinespace
        $\mathtt{Re750}$ & $2.706$ & $0.004$ \\
        $\mathtt{Re1750}$ & $2.406$ & $-0.068$ \\
        \addlinespace
        $\mathtt{m0(cuspy)}$ & $2.804$ & $-0.025$ \\
        \addlinespace
        $\mathtt{rc1560}$ & $2.620$ & $-0.024$ \\
        $\mathtt{rc2800}$ & $2.468$ & $-0.078$ \\
        \addlinespace
        $\mathtt{xc08}$ & $2.656$ & $-0.034$ \\
        $\mathtt{xc12}$ & $2.633$ & $-0.067$ \\
        \addlinespace
        $\mathtt{e02}$ & $2.582$ & $-0.042$ \\
        $\mathtt{e04}$ & $2.583$ & $-0.060$ \\
        \bottomrule
    \end{tabular}
\end{table}


\bsp	
\label{lastpage}
\end{document}